# New 111-type Semiconductor ReGaSi Follows 14e- Rules

Weiwei Xie*§, Lea Gustin §, Guang Bian ǂ

§ *Department of Chemistry, Louisiana State University, Baton Rouge, LA, USA, 70803*
ǂ *Department of Physics&Astronomy, University of Missouri, Columbia, MO, USA, 65201*

## Abstract

Electron-counting rules were applied to understand the stability, structural preference, and physical properties of metal disilicides. Following predictions made by 14 electron counting rules, the ordered semiconductor ReGaSi, the first ternary phase in this system, is proposed and successfully synthesized. It crystallizes with a primitive tetragonal structure (space group *P*4/*nmm*) closely related to that of $MoSi_2$-type $ReSi_2$, but with Ga and Si orderly distributed in the unit cell. The band structure, density of states, and crystal orbital calculations confirm the electron count hypothesis to predict new stable compounds. Calculations, based on 14 electrons per ReGaSi units, show a small indirect band gap of ~0.2 eV around Fermi level between full and empty electronic states. Additionally, first-principles calculations confirm the site preference of Ga and Si which is observed through the structural refinement. Experimental magnetic measurements verified the predicted non-magnetic properties of ReGaSi.

## Introduction

Semiconductors with various band structural features have enthused condensed matter and material engineering communities for many years due to their promising physical properties and practical applications.[1] Semiconductors with flat valence and conduction bands and large effective mass of carriers have demonstrated their potential as high-performance thermoelectric materials.[2] Certain *p*-type semiconductors with a good balance between optical transparency and electrical conductivity also present potential applications as transparent conductors in photovoltaics, such as K-doped $SrCu_2O_2$.[3,4] From the viewpoint of chemistry, what the electronic band structure represent to the extended solids is analogous to what the molecular orbital (MO) diagram does to a molecule. Similar to the MO diagram, the band structure plays critical roles in assisting scientists to interpret the structural stability and correlate crystal structures with physical properties such as electronic conductivity, optical/magnetic properties, and catalytic activity.[5] It is widely known that the structural stability of simple insulators or semiconductors can be explained by an intuitive bonding process in which an electropositive element donates electrons to an electronegative element and both achieve closed shell electron configurations. With increasing structural complexity, specific electron count rules such as 18e- count rules and Hume-Rothery rules were developed to explain the structural stability and led to the prediction of new compounds.[6,7,8] Among the most eminent examples are half-Heusler phases following 18e- count rules, for example, the noncentrosymmetric superconductor LuPtBi.[9,10] In addition to 18e- rules, a total valence electron count of 14e- per transition metal can be associated with the stability of some intermetallics such as topological insulating Nowotny Chimney Ladder phase, $Ru_2Sn_3$. Following the establishment of the 14e- rules, a handful of ternary compounds have been reported such as PbFCl-type pnictides of niobium, NbAsM (M= Ge and Si)[11] and ZrSSi-type TMM' (T= Hf and Zr; M= S, Se, and Te; M'= Si and Ge).[12]

In this paper, we study the band structure and molecular orbital diagram of semi-metallic $MoSi_2$ and propose that the 14e- rules may also work for $MoSi_2$-type compounds. Based on this, the new isoelectronic compound, ReGaSi, is designed and successfully synthesized. Moreover, the previous theoretical prediction by Mihalkovic et. al. of the primitive tetragonal structure of ReGaSi matches well with the experimental results in this paper.[13]

# Experiments Details

*Synthesis.* Polycrystalline ReGaSi was obtained by the high temperature metallic solution growth method using Ga as the self-flux. Different Re:Si:Ga compositions ranging from 1:1:48 to 1:2:47 were examined. The mixtures with elemental Re (99.9%; Alfa Aesar) Si (99.999%, Alfa Aesar) and Ga (99.999%, Alfa Aesar) were sealed into an evacuated silica tube ($< 10^{-4}$ Torr). They were heated to 950°C (1.5 °C/min) and allowed to anneal for 24 h, cooled down to 600°C at a rate of 10 °C/min and maintained at this temperature for 3 days. After the reaction, the silica tube was centrifuged to separate the sample from the unreacted gallium flux. All products appear to be stable upon exposure to air and moisture. Efforts were made to remove any remaining unreacted gallium using sonication or diluted HCl for extended periods of time. A small amount of Ga would still remain in the sample. The X-ray diffractions and physical properties measurements were performed on the specimen from the composition $ReSiGa_{48}$.

*Phase Analysis.* A Rigaku MiniFlex 600 powder X-ray diffractometer employing Cu radiation ($\lambda_{K\alpha}$=1.5406 Å) and a germanium monochromator was used to determine the phase identity and purity of the sample. The scattered intensity was recorded as a function of the Bragg angle (2θ) with a step of 0.005° 2θ in step scan mode, ranging from 5 to 90°. Phase identifications with calculated powder patterns based on the single crystal data were performed using Crystal Diffraction (Crystal Maker software). ReGaSi XRD pattern was slightly smoothed using Gaussian Convolution (parameter=2) on the Rigaku PDXL software for a better visualization.

*Structure Determination.* The crystal structure of ReGaSi was solved based on single crystal X-ray diffraction data. Small crystals (~0.01×0.01×0.01 mm$^3$) were picked from the sample and mounted on the tip of Kapton loop and loaded on a Bruker Apex II diffractometer with Mo radiation ($\lambda_{K\alpha}$=0.71073 Å). Reflections were collected with an exposure time of 10s per frame with 0.5° scans in ω in a 2θ range from 5 to 65°. Over 10 different crystals were tested to obtain reliable lattice parameters. The data acquisition, extraction of intensity and correction for Lorentz and polarization effects were performed using Bruker SMART software. The crystal structures were solved with the SHELXTL package, using direct methods and refined by full-matrix least-squares on $F^2$.[14] All crystal structure drawings were produced using the program *Vesta*.[15]

*Scanning Electron Microscopy (SEM).* Characterization was accomplished using a variable pressure scanning electron microscope (JSM-6610 LV) and Energy-Dispersive Spectroscopy (EDS) (Oxford Instruments X-ray analyzer). Samples were mounted on a thin layer of carbon tape prior to loading into the SEM chamber. Multiple points and areas were examined in each phase within multiple grains of a specimen. Compositional estimates were calculated using Oxford's SEM Quant software to correct intensities for matrix effects. The samples were examined at 20 kV and the spectra were collected for 100 seconds to get the chemical composition as accurate as possible. An Oxford Instruments Tetra backscattered electron (BSE) detector was used to image the samples using the BSE signal.

*Magnetic Measurements.* The magnetization measurements [zero-field cooling/field cooling (ZFC/FC) and hysteresis] were performed using a superconducting quantum interference device (SQUID) magnetometer MPMS XL-5 manufactured by Quantum Design, Inc. on pieces of ReGaSi cube-shaped single crystals. The SQUID was operated in the temperature range of 10-300 K in a magnetic field of 0.1 T. The samples were placed in plastic capsules for the measurements.

## Calculation Details

*Tight-Binding Linear Muffin-Tin Orbital -Atomic Spheres Approximation (TB-LMTO-ASA).*[16] The electronic and possible magnetic structures were calculated with TB-LMTO-ASA using the Stuttgart code. Exchange and correlation were treated by the local density approximation (LDA) and the local spin density approximation (LSDA).[17] The atomic spheres approximation method treats the space filled with overlapping Wigner-Seitz (WS) spheres.[18] The atoms in ReGaSi are homogeneously distributed in the space, so no empty sphere is necessary to conduct the calculation, and the WS sphere overlaps are limited to no larger than 18%. The symmetry of the potential is taken consideration into spherical inside each WS sphere, and the overlapping part has been corrected. The detailed WS radii are: 1.53 Å (Re), 1.52 Å (Ga), and 1.39 Å (Si). The basis set for the calculations included Re ($6s$, $6p$, $5d$), Ga ($4s$, $4p$) and Si ($3s$, $3p$) wavefunctions. The convergence criteria were set to $1 \times 10^{-5}$ eV. A mesh of 100 k points in the irreducible wedge of the first Brillouin zone was used to obtain all integrated values, including the density of states (DOS), Crystal Orbital Hamiltonian Population (COHP) curves, band structures, and molecular orbitals diagrams. [19]

*Vienna Ab-initio Simulation Package (VASP).*[20] The band structure calculation was completed with VASP, based on Density Functional Theory (DFT) using projector augmented-wave (PAW) pseudo potentials[21] that were adopted with the Perdew-Burke-Ernzerhof generalized gradient approximation (PBE-GGA).[22] Scalar relativistic effects were included. The energy cutoff was set at 500 eV. Reciprocal space integrations were completed over an 8×8×4 Monkhorst-Pack $k$-points mesh with the linear tetrahedron method. The calculated total energy is set to be converged to less than 0.1 meV per atom.

*WIEN2k*.[23] The electronic structures were also calculated using the full-potential linearized augmented plane wave and local orbitals basis WIEN2k code. Spin-orbit coupling (SOC) and relativistic effects on Re atoms were included. Experimental lattice parameters from single crystal diffraction refinement were used in the calculations. Because the calculations utilized in DFT-based and LMTO methods often underestimate band gaps, we introduced modified Becke-Johnson potential (mBJ) to correct it. [24]

## Results and Discussion

Our synthetic exploration was inspired by the geometric and electronic links among metal disilicides presented in Figure 1. Starting from Si itself, Si atoms form 4 bonds with the neighboring Si atoms, resulting in the diamond framework and the valence electron for $Si_2$ is 8e-. Upon reaction with electropositive elements such as Sr, tetragonal α-$SrSi_2$ phase would form. α-$SrSi_2$ with space group $I4_1/amd$ contains $(Si_2)^{2-}$, in which Si forms 3 bonds with the adjacent Si atoms along the *c*-axis.[25] The α-$SrSi_2$ phase has 10 valence electrons (2e-/Sr + 2×4e-/Si = 10e-).[26] The combination of Si and early transition metals leads to the formation of polar intermetallics, which can't be directly interpreted using Zintl-Klemm concepts which have been successfully applied to $MoSi_2$. Meanwhile, in the electronic structures of intermetallics, a feature related to the structural stability for a given chemical composition is the occurrence of a pseudogap around the Fermi level in the electronic density of states (DOS) curves. To examine the stability in $MoSi_2$ and $MoSi_2$-type $ReSi_2$, the DOS curves of $MoSi_2$ and $ReSi_2$ are generated. In both cases, the broad pseudogap, which is around Fermi level in $MoSi_2$ and ~2eV below Fermi level in $ReSi_2$, corresponds to 14e- valence electrons per transition metal atom. The analysis of the electron counts of two $MoSi_2$-type compounds, together with the pseudogap in the DOS curve in the 14e-/transition metal configuration, inspired us to search new ternary derivatives with 14e- counts.

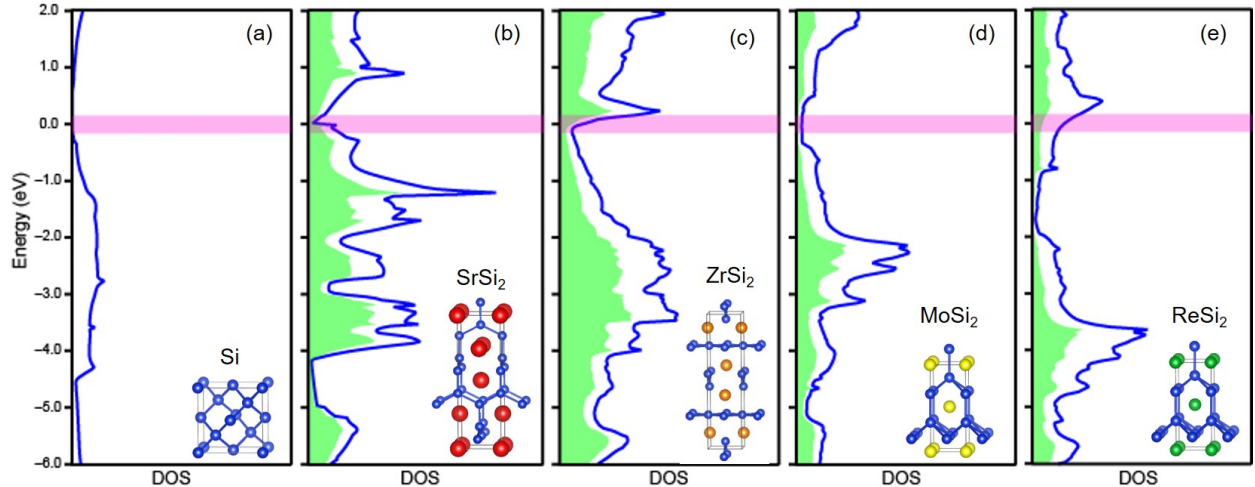

**Figure 1.** Crystal structures and density of states (DOS) curves of Si and metal disilicides with the emphasis on the Si-Si links.

*Phase Analysis:* The loading composition of ReSi$_2$Ga$_{47}$ yielded two mixed phases with different crystal shapes. One product formed brittle, rectangle-plate crystals that were proved to be MoSi$_2$-type ReSi$_2$ by single crystal X-ray diffraction and SEM-EDX, and the other one formed cube-shape crystals, which were ReGaSi by detailed characterizations. SEM images revealed that the ReGaSi exhibits a cube-like structure with a rough surface whereas ReSi$_2$ crystallizes as smooth rectangle platelets (Figure 2, right). The two compounds could also be distinguished through a regular optical microscope, allowing one to manually pick up ReSi$_2$ crystals from the mixture. The SEM-EDX analysis on the cube-like structure yielded the expected composition of Re$_{1.0(1)}$Ga$_{1.0(1)}$Si$_{1.0(1)}$ whereas the platelet samples were confirmed to be without Ga in them.

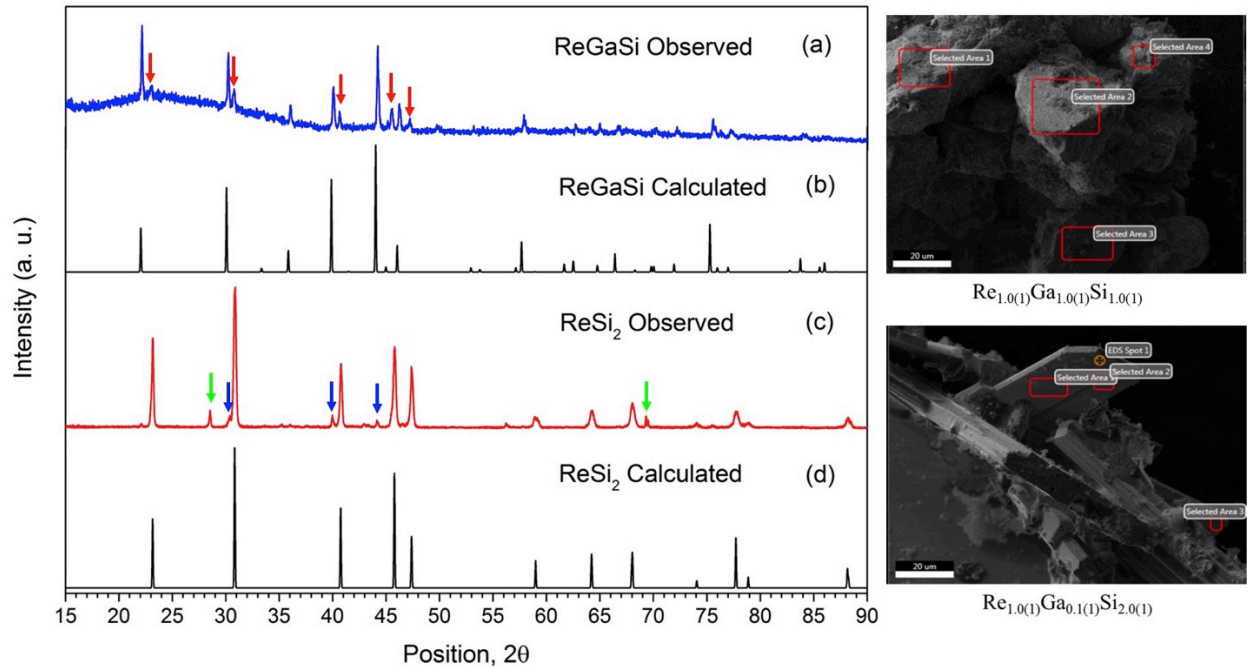

**Figure 2.** (Left) XRD patterns of (a) observed (sample ReSiGa$_{48}$) and (b) calculated ReGaSi, (c) observed (sample ReSi$_2$Ga$_{47}$) and (d) calculated ReSi$_2$. Impurity peaks from ReSi$_2$, ReGaSi and Ga flux were labeled with red, blue and green arrows, respectively. (Right) SEM images of (top) ReGaSi and (bottom) ReSi$_2$

The loading composition of ReSiGa$_{48}$ yielded a high purity phase of cube-shaped ReGaSi. Since the crystals were obtained from Ga-flux growth, there could always be some excessive Ga which could not be easily ground off for X-ray powder diffraction analysis, and this causes difficulties to perform the accurate resistivity measurements on the samples. But, according to X-ray powder diffraction, the patterns could be successfully indexed and matched with the calculated pattern obtained for single crystal diffraction experiments as shown in Figure 2(*a*) while the extra peak were explained by the presence of ReSi$_2$ in the sample. By contrast, the samples from ReSi$_2$Ga$_{47}$ yielded a majority phase of ReSi$_2$ according to the powder XRD pattern, see Figure 2(*c*). Additionally, higher amounts of remaining Ga flux were observed in the sample despite the different attempts to remove it.

*Crystal Structure and Site Preferences:* The structural determination was carried out on the new ReGaSi compound to determine accurate interatomic distances and coordination environments. Indexing of the PXRD pattern was consistent with a primitive tetragonal unit cell, which was further confirmed by the SXRD refinement. The unit cell parameters obtained from XRD appeared to be close to that of MoSi$_2$[27] which led us to use its structural model as the starting point for the full matrix least-squares refinement. The addition of Ga in the structure reduces crystal symmetry, from a body centered unit cell to a primitive cell which accommodates two distinct sites for Ga and Si, splitting the Si 4e site in MoSi$_2$ into two 2c sites. Primitive tetragonal space group *P*4/*nmm* (Pearson Symbol *tP*6) offered the best fit and therefore was used for the structure determination. Different refinement models were considered to verify the possibilities of disorders of site occupancies in ReGaSi phase.[28] At first, a random distribution of Ga and Si in the two sites was considered but the refinement did not lead to an acceptable fit. The ordering of the 2c sites (¼, ¼, z) was then investigated and the final refinement of the diffraction data led to a very good agreement with a configuration with Si sitting in the middle of the unit cell. Detailed electronic calculations further confirmed the site preference and will be discussed later on (see Band Structure section). Table 1 and 2 summarized the results of single crystal X-ray diffraction characterization of a specimen extracted from a single crystal sample of nominal composition ReGaSi. Corresponding anisotropic displacement parameters and significant interatomic distances are summarized in Tables S1 and S2. The structure of ReGaSi is shown in Figure 3 in comparison with that of ReSi$_2$.

**Table 1.** Crystallographic data for ReGaSi sample at 300 (2) K

| Refined Formula | ReGaSi |
|---|---|
| F.W. (g/mol); | 284.01 |
| Space group; Z | *P*4/*nmm* (No.129); 2 |
| Lattice Parameters | a= 3.194 (1) Å<br>c= 8.054 (2) Å |
| Volume (Å$^3$) | 82.18 (6) |
| Absorption Correction | Numerical |
| Extinction Coefficient | 0.047(9) |
| Θ range (deg) | 2.53 to 31.98 |
| *hkl* ranges | -4<=h, k<=4<br>-4<=l<=10 |
| No. reflections; $R_{int}$ | 245; 0.0297 |
| No. independent reflections | 108 |
| No. parameters | 11 |
| $R_1$; $wR_2$ (all *I*) | 0.0379;0.0887 |
| Goodness of fit | 1.249 |

| | Diffraction peak and hole (e⁻/Å³) | 3.108; -4.604 |
|---|---|---|

**Table 2.** Atomic coordinates and equivalent isotropic displacement parameters of ReGaSi systems ($U_{eq}$ is defined as one-third of the trace of the orthogonalized $U^{ij}$ tensor (Å²)).

| Atom | Wyck. | Occ. | x | y | z | $U_{eq}$ |
|---|---|---|---|---|---|---|
| Re | 2c | 1 | ¾ | ¾ | 0.7306(1) | 0.0053(5) |
| Ga | 2c | 1 | ¼ | ¼ | 0.9146(3) | 0.0081(7) |
| Si | 2c | 1 | ¾ | ¾ | 0.4152(8) | 0.0060(16) |

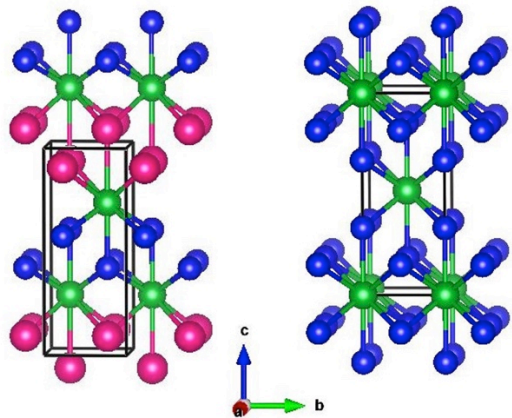

**Figure 3.** Structural comparison between ReGaSi (left) and ReSi₂ (right) [Green: Re; Blue: Si; Pink: Ga]

*Structural Comparison with MoSi₂-type ReSi₂ and other Primitive Tetragonal 111 Phases:* To get a better understanding of the new ReGaSi structure, one can put it in context with all other primitive tetragonal 111 phases. The four most common structure types reported up to date are presented in Figure 4.

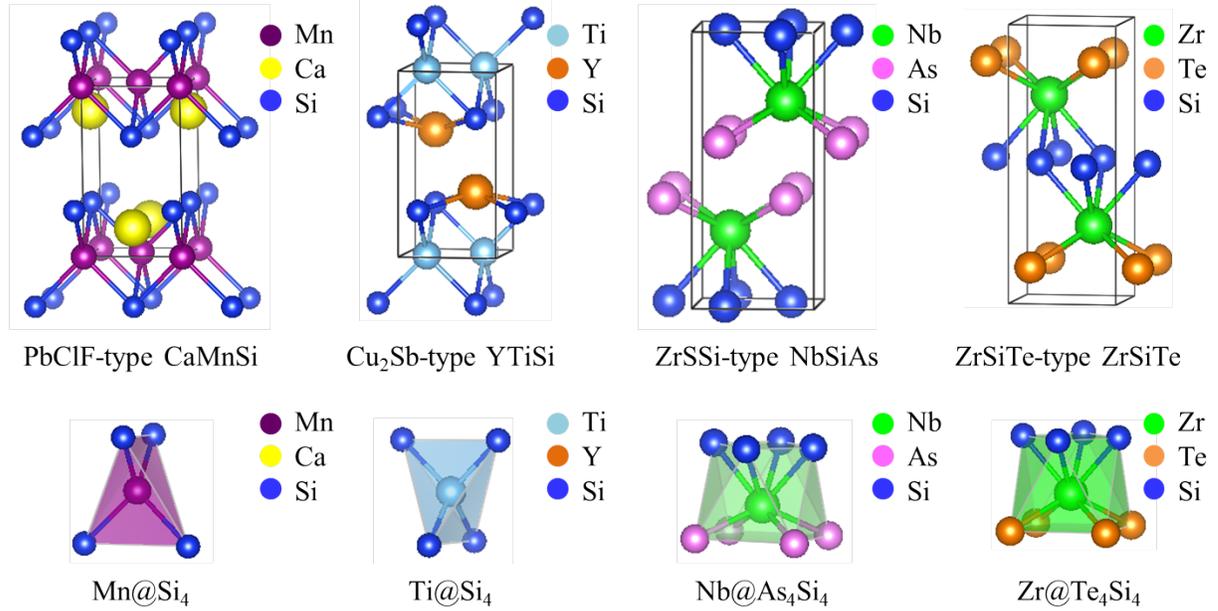

**Figure 4.** Crystal structures of the archetypal primitive tetragonal 111 compounds CaMnSi, YTiSi, NbAsSi, and ZrTeSi. The top panel illustrates the stacking along the c axis; the bottom panel demonstrates the clustering environments.

PbClF- and $Cu_2Sb$-types are both buckled tetragonal structures. The PbClF-type CaMnSi shown in Figure 4 can be written in the ionic formula as $Ca^{2+}Mn^{2+}Si^{4-}$ with Mn atoms centering in the Si tetrahedral clusters.[29] The Si-Ca-Mn stacking along *c*-axis is an indication of a strong charge transfer among the atoms consistent with the semiconducting property of the compound. $Cu_2Sb$-type YTiSi contains similar Ti-centered Si tetrahedral clusters but the electronegativity difference between the atoms is smaller than that in CaMnSi.[30] As a result, YTiSi shows less layered structural features and a more metallic behavior. The ZrSSi- and ZrSiTe-type structures for NbSiAs and ZrSiTe, respectively, display unbuckled tetragonal structures with Zr-centered anti-square prisms.[13] With its $Re@Ga_5Si_5$ clusters, the primitive tetragonal structure of the new ReGaSi clearly stands out among the four 111-type compounds presented above. The multiple Re-Ga and Re-Si contacts indicate the strong interactions within the polyhedral clusters, thus motivating us to investigate its electronic structure and possible properties in details.

*The 14e- rule and Molecular Perspectives on ReGaSi:* Before embarking on the electronic structure of ReGaSi calculated by first-principles methods, a schematic picture on the basis of a molecular perspective of $MoSi_2$, $ReSi_2$ and $Re_2Ga_2Si_2$ (formula unit), which corresponds to our original motivation of using the electron counting rules to predict new compounds, is discussed below. [31,6] For $MoSi_2$, the detailed orbital analysis on Γ, Z, X and P and N-points shows there are 6 orbitals distributed in energy up to Fermi level, which primarily originate from 1×s, 1×(s-p) hybrids, 3× (d-p) hybrids, and 1×d orbitals in Figure 5.

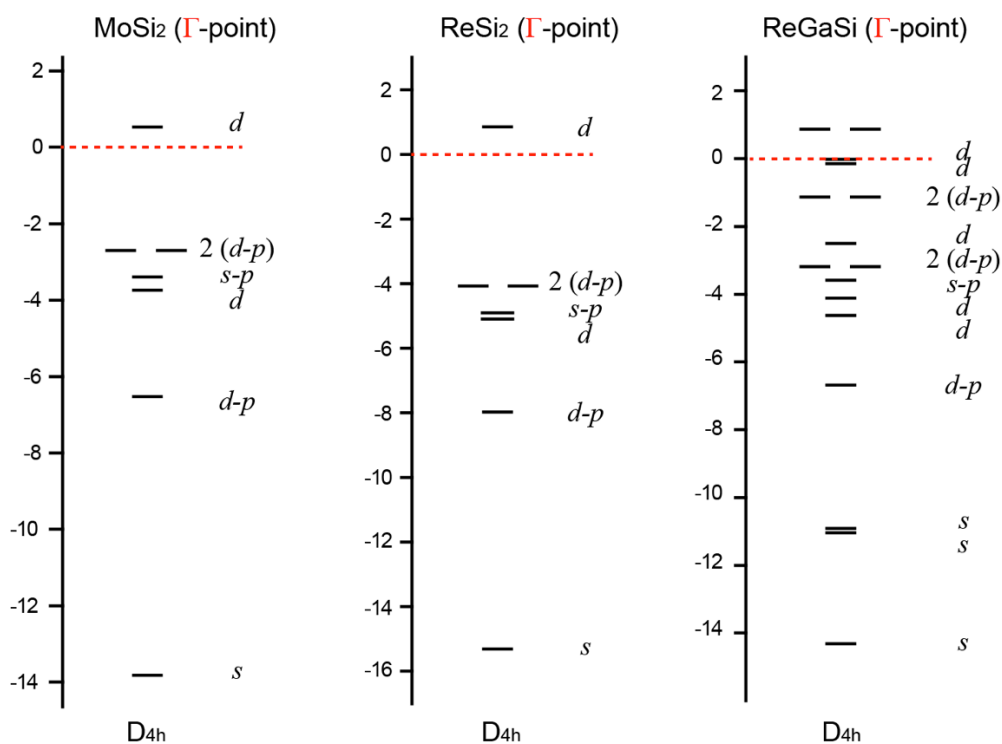

**Figure 5.** Molecular energy level diagrams for $MoSi_2$, $ReSi_2$, and ReGaSi generated by LMTO method.

In this scheme, 14 electrons are needed to achieve the 7 occupied orbitals per $MoSi_2$ unit. The electron counting yields 6 and 4 electrons for Mo and Si atoms, respectively, which results in a total of 14 valence electrons per $MoSi_2$ unit. Thus, the seven bonding orbitals in the cluster should be fully occupied and one can expect the Fermi level to be located at a gap or pseudogap in the DOS. [32] Similarly, the orbital analysis of $ReSi_2$ also indicates the scheme of 14 electrons per formula unit (f.u.) is preferred from the perspective of electronic stability. However, Re has one extra valence electron than Mo, thus the Fermi level of $ReSi_2$ corresponds to 15e- per formula rather than 14e-. Coming to the ReGaSi, which has the same point group as $MoSi_2$ and $ReSi_2$ but doubles the primitive unit cell, the orbitals below the Fermi level are primarily from 3×$s$, 1×($s$-$p$) hybrids, 5×($d$-$p$) hybrids, and 5×$d$ orbitals. A total of 28e- per $Re_2Ga_2Si_2$ (14e- per ReGaSi) is needed to fill up these orbitals and make the Fermi level in the gap or pseudogap in the DOS. Generally, the 14e- rule can be extended to transition metal complex, especially $TM_2$ ($T$= transition metals; $M$ = Group III and IV elements) with a similar routing. Moreover, tetragonal $MoSi_2$, ReGaSi, and twisted Chimney Ladders phases confirm the 14e- rule among the intermetallics and indicate the structure type does not limit the rules. [32]

*Electronic Structure and Isolated-Clustering Properties of ReGaSi:* TB-LMTO-ASA calculations were carried out to evaluate and analyze the electronic DOS and bonding interactions, which can help get a better understanding of the electronic influence on the atomic interactions. As expected from the electron counting described previously, the partial DOS curves (Figure 6a) emphasize the contributions from the valence orbitals of each atom and show the Fermi level ($E_F$) lies in the band gap.

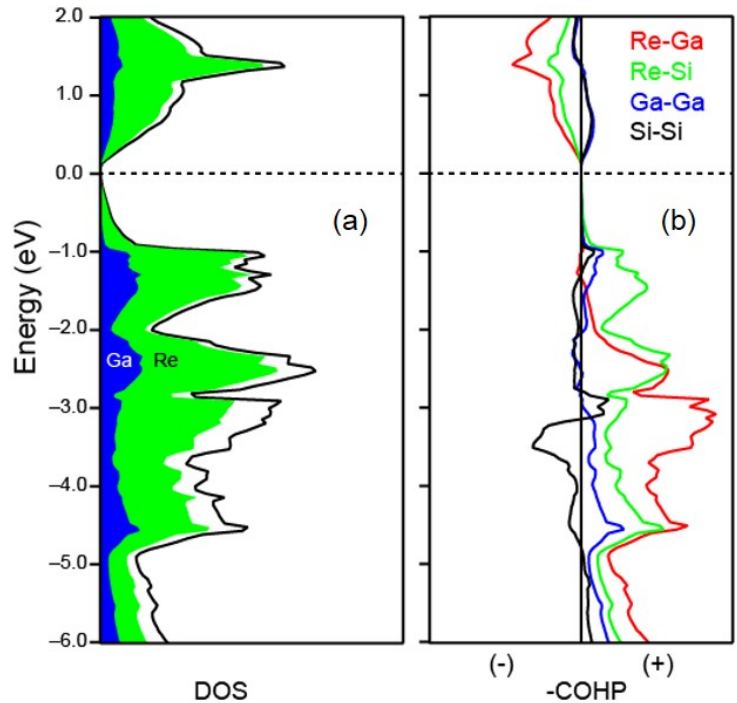

**Figure 6.** (a) Partial DOS curves of ReGaSi emphasizing Re contribution in green, Ga contribution in blue and Si contribution in white. (b) −COHP curves of ReGaSi obtained from non-spin polarization (LDA; + is bonding/− is antibonding).

The small band gap around the Fermi level was determined based on the band structure calculation with 14 electrons per ReGaSi unit, and the gap size was found to be ~0.2 eV. The band gap calculation was also calculated with different methods (LDA+SOC+mBJ by WIEN2k: ~0.2 eV; LMTO-ASA: ~0.1eV; LDA+SOC by VASP: ~0.17eV). As can be seen on the partial DOS curves, Re 5$d$ orbitals contribute dominantly to the DOS curve between −5 and +2 eV, whereas, the valence bands (4$s$ and 4$p$) of Ga homogenously distribute in the DOS. On the other hand, the 3$s$ and 3$p$ orbitals of Si mainly distribute below -3 eV in the DOS.

According to the corresponding –COHP curves (Figure 6b), the wave functions of bonding orbitals are dramatically influenced by the strong Re-Ga and Re-Si interactions. The band gap at Fermi level is associated with the optimization of bonding and antibonding interactions in this structure. According to the LDA-DOS curves and the evaluation of the Stoner condition, ReGaSi is susceptible to be non-magnetic.[33] To confirm the non-magnetic property, spin polarization was applied *via* the local spin density approximation (LSDA) to split the DOS curves into the spin-up and spin-down parts. The total DOS curve with both spin-up and spin-down in LSDA is identical with the DOS curve obtained by LDA. Furthermore, the integration of the spin-up and spin-down DOS curves yields a total magnetic moment of 0 $\mu_B$ per ReGaSi, indicative of the nonmagnetic nature of this compound. In spite of the electronic features of ReGaSi, an intriguing character in –COHP is the abnormally weak Si-Si and Ga-Ga interactions. The Si-Si distance is ~2.65 Å, slightly longer than the Si-Si distance in TiSi$_2$ (2.55 Å) and close to the Si-Si distance in Chimney Ladders phase, MnSi$_{1.75-\delta}$ (2.66 Å).[34] In addition, the Ga-Ga distance (2.64 Å) is similar to that in the elemental state of Ga. The weak Ga-Ga and Si-Si interactions along with strong Re-Ga and Re-Si interactions indicate the isolated molecular clustering property of ReGaSi. As a result, one can view the ReGaSi structure as repetitive Re@Ga$_5$Si$_5$ clusters.

To gain more accurate band information in ReGaSi, density functional theory based calculations were performed by using VASP and the results are shown in Figure 7. In a good agreement with the band structure derived from the LMTO calculation, the appearance of band gap around the Fermi level can be simply understood by the fact that it is a charge-balanced compound; we therefore expect a semiconducting behavior of ReGaSi. Owing to the strong atomic spin-orbit coupling of Re atoms, the band degeneracy, especially at Γ-point, are lift, which are clearly shown in Figure 7. The LDA band structure reveals that the Re 5$d$ bands close to the Fermi level are very flat, nearly dispersionless along Γ−Z. To examine possible electronic influences on the site preferences of Si and Ga atoms in ReGaSi and confirm the site preference determined in the structural refinement, VASP calculations were performed to evaluate the total energies and electronic structures of ReGaSi in two different crystal structures (models I and II).

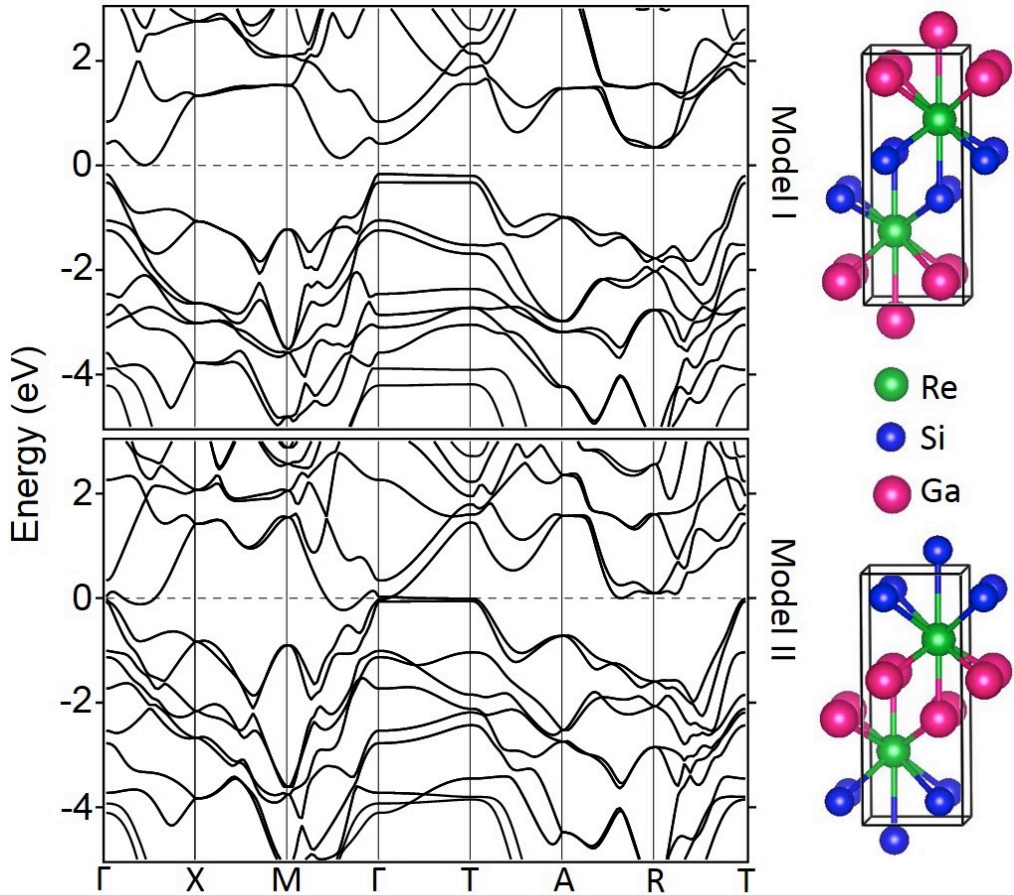

**Figure 7.** Top: Band structure of ReGaSi-Model I generated by VASP with spin-orbital coupling. Bottom: Band structure of ReGaSi-Model II generated by VASP with spin-orbital coupling.

In Model I, the same as the refined structure, Ga occupies the outside 2c site of the structure while Si is located at the center of the unit cell (¼, ¼, 0.4152). In Model II, the sites of Ga and Si are interchanged compare to Model I. According to the calculated total energies, the experimental result (Model I) with Ga fully at the outside 2c sites (¼, ¼, 0.9146) clearly gives a lower energy (0.989eV/f.u. difference in total energy). To gain further understanding of the site preference, the band structures of two models are generated using VASP, see Figure 7. Compared to the indirect semiconducing band gap of size ~0.2 eV in Model I, the band structure of Model II shows a semimetallic behavior with a small overlap between the conduction and valence bands around Fermi level, making Model II energetically less favorable compared to Model I. This confirms the structure obtained from the diffraction data.

*Diamagnetic Properties of ReGaSi:* To prove the magnetic properties predicted by theoretical calculations, magnetic measurements were performed on the ReGaSi crystals. The relatively small temperature-independent molar magnetic susceptibility, presented in Figure 8(a), is dominated by core diamagnetism; the magnetic susceptibility of the material is around $-6\times10^{-6}$ emu/mol f. u.. This indicates that the paramagnetic contribution of conduction electrons to the observed susceptibility is small, which is also an indirect evidence to support the charge-balanced characteristics of ReGaSi. Because of the Ga-flux in the sample, the resistivity measurement cannot be conducted accurately to show the semiconducting or semimetallic behavior. The magnetic hysteresis measurements in Figure 8(b) show the field-dependent magnetization at 10 K is diamagnetic, which is in agreement with the electronic structure calculations.

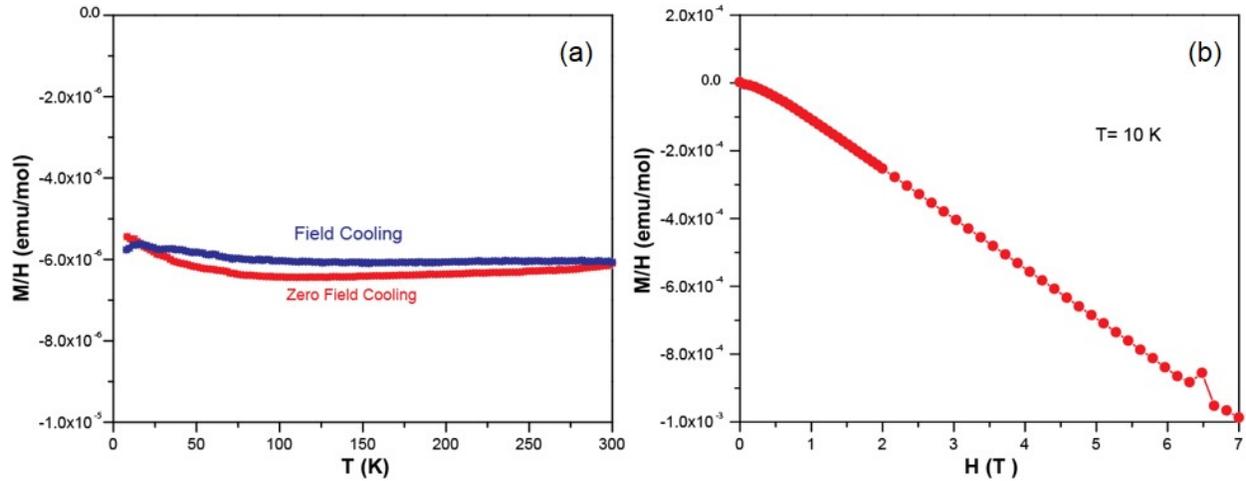

**Figure 8.** Magnetic measurements for $MoSi_2$-type ReGaSi indicating the weak diamagnetic property.(a) Temperature-dependent magnetic curve measured with 0.1 Tesla applied field under the zero-field cooling and field cooling conditions. (b) Magnetic hysteresis curves at 10 K.

## Conclusions

The first ternary phase in Re-Ga-Si systems, ReGaSi was synthesized, characterized, and analyzed by XRD, magnetic measurements and electronic structure calculations. The relationship between crystal and electronic structures among different metal disicilides offers new insights into the role played by electron counts in the structural stability of intermetallic compounds. The molecular orbital analysis indicates the new 111-type tetragonal ReGaSi follows the 14e- rule. Moreover, first-principles electronic structure calculations reveal the site preference and a semiconducting indirect band gap in the electronic structure of ReGaSi. The strong Re-Ga and Re-Si interactions and weak atomic bonding of Si-Si and Ga-Ga result in the isolated clustering molecule-like properties of ReGaSi.

## Associated Content:

Supporting Information Available:

Anisotropic displacement parameters and interatomic distances (PDF)

X-ray crystallographic file for compound ReGaSi (CIF)

## Acknowledgements

We deeply appreciate the financial support from Louisiana State University (startup funding) and the Shared Instrument Facility (SIF) at Louisiana State University for the SEM-EDS and SQUID magnetic measurements. W. Xie thanks Prof. Robert Cava at Princeton University for offering the access to the VASP program and giving comments on English writing.

**For Table of Contents Only**

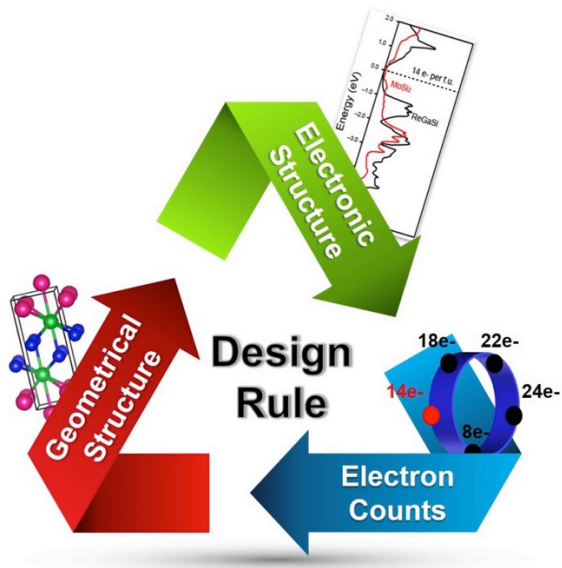

**Synopsis**

The first ternary phase in Re-Ga-Si system, ReGaSi, was designed based on electron count rules and confirmed by experimental results. ReGaSi adopts a framework similar to that of tetragonal $MoSi_2$, with Ga and Si orderly distributed on the two sites, lowering the symmetry from body-centered to primitive. The electronic structure calculations on ReGaSi show a small indirect band gap, ~0.2 eV, around Fermi level.